\documentclass[aip,preprint]{revtex4}
\usepackage{graphicx}% Include figure files
\usepackage{dcolumn}% Align table columns on decimal point
\usepackage{bm}% bold math
\usepackage[english]{babel}

\begin{document}
\title{Double proximity effect in hybrid planar Superconductor-(Normal metal/Ferromagnet)-Superconductor structures}
\author{T.~E.~Golikova$^{a}$\email{golt2@list.ru},
F.~H\"ubler$^{b}$, D.~Beckmann$^{b}$, I.~E.~Batov$^{a}$, T.~Yu.~Karminskaya$^{c}$, M.~Yu.~Kupriyanov$^{c}$, A.~A.~Golubov$^{d}$, and V.~V.~Ryazanov$^{a}$}
\affiliation{$^{a}$Institute of Solid State Physics RAS, 142432
Chernogolovka, Moscow district, Russia\\$^{b}$Institute of Nanotechnology, Karlsruhe
Institute of Technology, 76021 Karlsruhe, Germany
\\$^{c}$Lomonosov Moscow State University, Skobeltsyn Institute of Nuclear Physics, Moscow 119991, Russian Federation\\$^{d}$Faculty of Science and Technology, University  of Twente, 7500 AE Enschede, The Netherlands}
\begin{abstract}
We have investigated the differential resistance of hybrid planar
Al-(Cu/Fe)-Al submicron bridges at low temperatures and in weak magnetic
fields. The structure consists of Cu/Fe-bilayer forming a bridge between two superconducting Al-electrodes. In superconducting
state of Al-electrodes, we have observed a double-peak peculiarity
in differential resistance of the S-(N/F)-S structures at a bias voltage
corresponding to the minigap. We claim that this effect (the doubling
of the minigap) is due to an electron spin polarization in the normal metal which is
induced by the ferromagnet. We have demonstrated that the double-peak
peculiarity is converted to a single peak at a coercive applied field
corresponding to zero magnetization of the Fe-layer.
\end{abstract}

\maketitle

\indent In superconductor-normal metal
(SN) bilayers the superconducting proximity effect is responsible for the modification of the electron density of
states (DOS) and the appearence of a minigap ${\varepsilon_g}$ in the normal metal \cite%
{Gol_Kupr_1988}. Thereby a normal metal region close to the SN-interface
behaves as a genuine superconductor, i.e. there is an energy range (-${\varepsilon_g}$%
,+${\varepsilon_g}$) around the Fermi energy in which there are no available states
for normal quasiparticles. This theoretical statement \cite{Gol_Kupr_1988}
was proved reliably in recent measurements of the local DOS \cite%
{Gueron,Courtois,Esteve}. A minigap peculiarity becomes apparent in the
differential conductance (resistance) spectra of SNS junctions side by side
with the superconducting gap peculiarity of superconducting electrodes \cite%
{Cuevas,Giazotto}. In the case of a superconductor-ferromagnet (SF) bilayer the
ferromagnetic exchange splitting of the spin sub-bands results in an energy shift of the corresponding minigap, which is asymmetric for the majority and the minority spin sub-bands \cite{SFIFS}, i.e. one can distinguish two minigap
peculiarities in SF-DOS spectra. However, even in the case of diluted ferromagnets the exchange field ${E_{ex}}$ is very large, so it is difficult to observe the minigap splitting on well-known proximity SF-systems like Nb-CuNi \cite{Oboznov} and
Nb-PdNi \cite{Kontos}. In Ref. \cite{Yip}, Yip first proposed to modify the
DOS in SN-proximity system by applying a magnetic ``Zeeman'' field \textsl{h}. Unfortunately, it is difficult to use applied magnetic fields in real SN-experiments due to the ``orbital'' suppression of the superconducting
electrodes. Recently, authors of Ref. \cite{Karm1} have proposed to induce a weak ``exchange field'', ${h_{ef}}$, via diffusion of spin polarized electrons from F to N metal in NF bilayers,
i.e. by using a complex NF bilayer as weak link in a S-(N/F)-S structure.
In this case an ``effective'' exchange field ${h_{ef}}$ which is
induced in the N layer is much smaller than the intrinsic exchange field ${E_{ex}}$ of the neighboring F-layer.
\begin{figure}[tbp]
\centering
\includegraphics[width=0.5\textwidth]{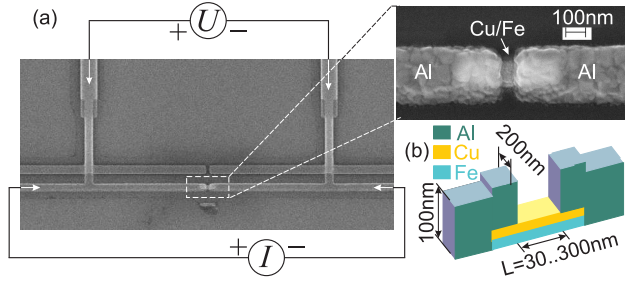}
\caption{(color online)(a) SEM image of the Al-(Cu/Fe)-Al junction together with
the measurement scheme. Inset shows the central part of the junction. (b)
The schematic sketch of the sample with geometrical dimensions }
\label{scheme}
\end{figure}

In this work, we report on the first experimental observation of the
``minigap doubling'' in the SNF banks of S-(N/F)-S submicron-size bridges
schematically shown in Fig.\ref{scheme}(b). The minigap doubling has been observed as
double peak peculiarity in the differential resistance of planar Al-(Cu/Fe)-Al
junctions fabricated by \textit{e}-beam lithography and the shadow
evaporation techniques.

Figure \ref{scheme}(a) shows a scanning electron microscopy (SEM) image of one of our
samples together with the measurement scheme. The submicron-scale planar
junctions were fabricated by means of the electron beam lithography and \textit{in situ} shadow evaporation. First, a thin (10-15 nm) iron layer is deposited onto the oxidized silicon substrate, followed by the  deposition of a 60 nm thick copper layer, so that in combination the NF bilayer bridge (0.2$\times$(0.3$\div$0.6) $\mu m^2$) is formed. Subsequently, a thick aluminum layer of around 100 nm is evaporated at a second angle in order to form the superconducting leads.
We fabricated samples with different separation length \textit{L} between
the superconducting electrodes, ranging from 30 nm to 300 nm. All transport
measurements were performed using standard four-terminal method. As the
specific resistance of the copper film ($\rho_{N}$=4.5 $\mu\Omega$$\times$cm)
is much smaller than the one of the iron film ($\rho_{F}$=70 $\mu\Omega$$\times$cm), the main part of the current flows through the copper layer. The
measurements at temperatures down to 0.3 K were performed in a shielded
cryostat equipped with a superconducting solenoid. Two stages of RC filters
were incorporated into the measurement system to eliminate the electrical
noise.

\begin{figure}[tbp]
\centering
\includegraphics[width=0.47\textwidth]{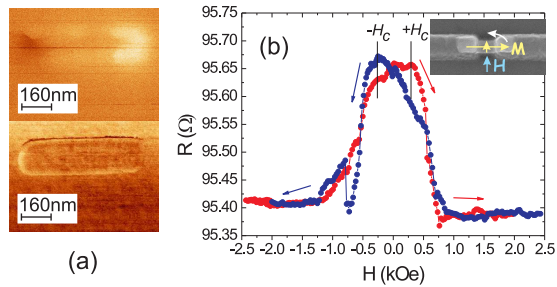}
\caption{(color online) (a) The magnetic (top) and topographical (bottom)
images of an iron bar with thickness of 10 nm. The images were edited by
using WSxM \protect\cite{wsxm}. (b) Resistance \textsl{R} of the
Al-Fe-Al junction vs. the external magnetic field \textit{H} at the
temperature 4.2 K, Inset: the SEM image of the sample with schematic view of
direction of the external magnetic field \textit{H} and magnetization
\textit{M} of iron layer.}
\label{magn}
\end{figure}

In order to check that the iron layer forms a single-domain magnetized along the S-(N/F)-S junction, reference structures with the same geometry and structure as the N/F-bilayers, but only with the ferromagnetic layer, were fabricated and subsequently investigated by means of magnetic-force-microscopy imaging (MFM).
Fig.\ref{magn}(a) shows a MFM image of the iron bar at
zero magnetic field together with the topographical  image (AFM). The picture of
magnetic poles is similar to the MFM images of iron nanostrips published in Ref.\cite{Hanson}. According to this work we dealt with practically uniform magnetized
structure. The main magnetization is directed along the long axis of
rectangle but diverges from a dipolar configuration at the corners \cite
{magnSim}. Non-local spin-valve experiments on similar submicron iron structures
indicate single-domain behavior, with coercive fields of about 200-500
Oe for magnetic fields applied along the element\cite{BeckmannCAR}. To estimate coercive field of the iron bar S-F-S (Al-Fe-Al)
bridges with the same geometry but without the Cu layer were prepared. We have
measured the magnetoresistance of the S-F-S bridge at \textit{T}=4.2 K using in-plane
magnetic field perpendicular to the Fe-bar easy-axis (Fig \ref{magn}(b)). The coercive
field $H_{c}$ (about 300 Oe) was determined from the maximum value of the resistance due to anisotropic magnetoresistance (AMR) effect (see, for example Ref. \cite{AMRexp,AMRt}). The observation of a finite coercive field suggests that the magnetization
configuration deviates from the single-domain structure during magnetization reversal.

\begin{figure}[tbp]
\centering
\includegraphics[width=0.47\textwidth] {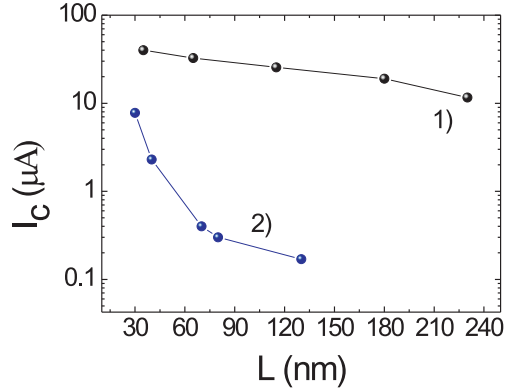}
\caption{(color online) Dependences of the critical current \textit{I$_{c}$}
on the sample length \textit{L} for (1) Al-Cu-Al and (2) Al-(Cu/Fe)-Al nanobridges at \textit{T}=0.4 K.}
\label{Ic}
\end{figure}

Resistive and Josephson characteristics of the planar junctions depend
strongly on the spacing, \textit{L}, between the aluminum electrodes as well as
on the total length of Cu/Fe-bilayers that were partly overlapped by the
electrodes. The characteristics and their discussion will be given in detail
later \cite{SNSTG}. The Josephson supercurrent was observed in structures with \textit{L} from 30 nm up to 130 nm. It
is important to note that the coherent Josephson transport was suppressed
significantly by addition of the extra ferromagnetic layer. Fig.\ref{Ic} presents
the dependence of the critical current \textit{I$_{c}$} vs. \textit{L}
for Al-(Cu/Fe)-Al junctions shown in Fig.\ref{scheme} in comparison with \textit{I$_{c}$%
(L)}-dependence for control Al-Cu-Al structures fabricated by the same
procedure but without the additional Fe-layer. The critical currents of S-(N/F)-S junctions are much smaller than that for S-N-S junctions.

The strong suppression of the superconducting proximity effect in the
N-channel of the S-(N/F)-S junction is due to the penetration of
spin-polarized electrons in the copper layer from the single-domain iron strip
underneath. We suppose that penetration of
spin-polarized electrons through the Fe/Cu-interface provides the copper
layer with an uniform spin polarization because of the small Cu-layer thickness
(of 60 nm). The data of Ref.\cite{Kimura} indicate that the spin-diffusion
length in Cu is as large as 1 $\mu$m at 1 K, that is larger than the bridge
sizes.

\begin{figure}[tbp]
\centering
\includegraphics[width=0.5\textwidth]{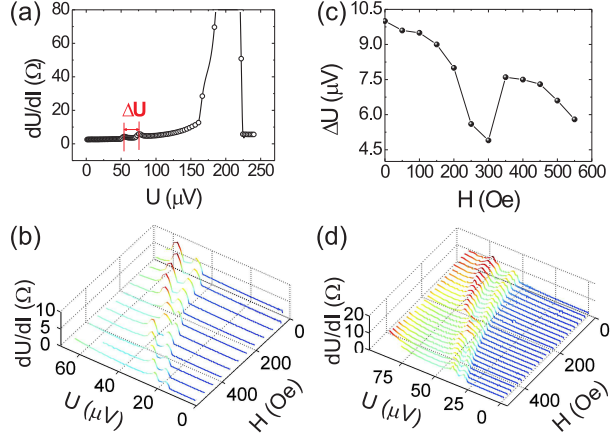}
\caption{(color online)(a)Differential resistance \textit{dU/dI} versus
voltage \textit{U} of Al-(Cu/Fe)-Al nanobridge S1 with length \textit{L}=130
nm at the temperature \textit{T}=0.4 K, the double peculiarity is signed with red lines,
\textit{$\Delta$U} - is the separation between peaks. (b) Differential resistance \textit{dU/dI} versus
voltage \textit{U} and external magnetic field \textit{H} of the sample S2;
(c) The distance between two peaks on the dependence (b) versus external
magnetic field \textit{H}; (d) Differential resistance \textit{dU/dI} versus
voltage \textit{U} and external magnetic field \textit{H} of the sample S3. Double peak becomes to single at the $H=H_{c}\approx300$ Oe. }
\label{result}
\end{figure}

To detect DOS peculiarities of the novel double-proximity
structures we measured differential current-voltage characteristics
by current-driven lock-in technique as well as the dc current-voltage
characteristics of the structures.
Fig. \ref{result}(a) demonstrates the differential resistance vs bias voltage for Al-(Cu/Fe)-Al junction (S1) with
the space \textit{L}=130 nm between superconducting electrodes at \textit{T}=0.4 K.
The curve is symmetric with respect to the zero bias voltage, therefore only positive voltage values are shown. There are two types of peculiarities on the \textit{dU/dI(U)}
dependence. First one corresponds to the superconducting gap of aluminum $%
\Delta$=180 $\mu$eV and the second one is a double-peak peculiarity at the
subgap energy $\varepsilon$$\approx$60 $\mu$eV which is much smaller than $\Delta$.

We suppose that the double-peak peculiarity in S-(N/F)-S transport is due to the
presence of two spin-dependent minigaps in the normal metal interlayer of SNF
trilayered electrodes. The most easy way to check this idea
with the spin-dependent minigap origin is to change the uniform state of the
ferromagnet layer magnetization. 
The differential resistance of the S-(N/F)-S
samples was measured in presence of magnetic field \textit{H} which
increases from zero by small steps (see Fig.\ref{result}(b)). Magnetic field was applied in plane of
the sample perpendicular to the bridge, as it was for S-F-S structures shown
in figure \ref{magn}(b) inset. One can see that at around 300 Oe the
separation between two peaks of the double-peak peculiarity decreases
significantly and then goes practically to the initial value at further increase
of the magnetic field.
The magnetic field of about 300 Oe coincides with the coercive field of the S-F-S reference structures. While the precise magnetization state at this field is unknown, a strongly inhomogeneous state can be expected, effectively reducing the induced exchange splitting in the N layer.
The dependence of the double-peak splitting \textit{$\Delta$U} vs. applied
magnetic field \textit{H} is shown in figure \ref{result}(c). Moreover it was observed for
some samples that the double-peak peculiarity joined to single peak at about 300 Oe (Figure \ref{result}(d)). The position of the
peculiarity is shifted to low voltages with increasing magnetic field
because of the suppression of superconductivity in both the aluminum electrodes
and the proximity region.

Below we briefly describe calculations of the minigap splitting in a SNF-NF
structure with trilayered electrode, i.e. appearance of two minigaps for
majority and minority spin systems in N-layer due to proximity effect from
the superconductor and magnetic proximity effect due to a contact with a
ferromagnet.

For simplicity we shall discuss the case when F and N films are thin
compared to the coherence lengths in these metals. Such simplification
allows to obtain simple solution for the gap splitting but does not change
our conclusions qualitatively. We assume that the dirty limit conditions are
fulfilled in the investigated structure,
therefore one can use the quasiclassical Usadel equations for
Green's functions which in $\theta $-parametrization have
the form:
\begin{equation}
\frac{\xi_{F,N}^{2}}{\widetilde{\Omega}}\left\{ \frac{\partial^{2}}{\partial x^{2}}\theta_{F,N}+\frac{\partial^{2}}{\partial y^{2}}\theta_{F,N}\right\} -\sin\theta_{F,N}=0.\label{Ue}\end{equation}
Here $\widetilde{\Omega }=\Omega+ih$, $h=E_{ex}/\pi T_{C}$, ${\Omega}=(2n+1)T/T_c$ are normalized Matsubara frequencies,
$E_{ex}$ is exchange field which vanishes in N metal, and $x(y)$-axes are parallel
(perpendicular) to the FN interface with the origin at the
boundary between the SNF trilayer and NF bilayer. Equations (\ref{Ue})
should be supplemented by the boundary conditions Ref.%
\cite{KL}
\begin{equation}
\gamma _{BN}\xi _{N}\frac{\partial }{\partial y}\theta _{N}=-\sin (\theta
_{S}-\theta _{N})
\end{equation}%
at the SN interface with $\gamma _{BN}=R_{B}/\rho _{N}\xi _{N}$ and
\begin{equation}
\xi _{N}\frac{\partial }{\partial y}\theta _{N}=\gamma \xi _{F}\frac{%
\partial }{\partial y}\theta _{F}, \theta _{N}=\theta _{F}
\label{eq:bc}
\end{equation}%
at the NF interface with $\gamma =\rho _{N}\xi _{N}/\rho_{F}\xi _{F}$ (we assumed that FN interface is transparent).
Here $\sin \theta _{S}=\Delta /\sqrt{\Omega ^{2}+\Delta ^{2}}$
and $\Delta $ is bulk pair potential of a superconductor. $R_{B}$ is the specific resistance of SN interface, $\rho _{S,F,N,}$
and $\xi _{S,F,N}$ are the resistivities and the coherence lengths of the S, F
and N layers. We assume that $\gamma _{BN}\gg \min \left(
1, \rho_S \xi_S/\rho_N \xi_N \right)$, so that suppression of superconductivity in S
electrode is negligibly small. At the free interfaces derivatives of $%
\theta $-functions are zero in the direction of the interface normal.

\begin{figure}
\centering \includegraphics[width=0.4\textwidth]{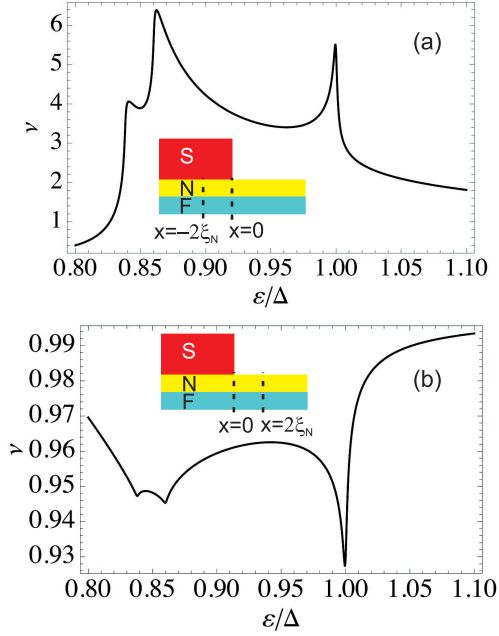}
\caption{N-layer DOS of the SNF-NF structure versus normalized energy $\varepsilon/\Delta$
at $\gamma_{BM}$=0.3, $h=0.05$, $\gamma=1$ for
(a) $x=-2\xi_N$ and (b) $x=2\xi_N$}
\label{th}
\end{figure}

The problem (\ref{Ue})-(\ref{eq:bc}) is reduced to one-dimensional equations for
Green' functions in the NF bilayer under S electrode $\theta _{-}$ (for $x<0$%
) and Green's functions for free FN bilayer $\theta _{+}$ (for $x>0$):%
\begin{equation}
\begin{array}{c}
\eta ^{2}\frac{\partial ^{2}}{\partial x^{2}}\theta _{-}-\sin \left( \theta
_{-}-\theta _{-\infty }\right) =0, \\
\mu ^{2}\frac{\partial ^{2}}{\partial x^{2}}\theta _{+}-\sin \theta _{+}=0,%
\end{array}%
\end{equation}%
where
\[
\eta ^{2}=\frac{\gamma _{BM}\left( \gamma k\xi _{F}^{2}+\xi _{N}^{2}\right)
\cos \theta _{-\infty }}{\gamma _{BM}(\gamma k\widetilde{\Omega }+\Omega
)+\cos \theta _{S}},\,\mu ^{2}=\frac{\gamma k\xi_{F} ^{2}+\xi _{N}^{2}}{\gamma k%
\widetilde{\Omega }+\Omega },
\]
and $k=d_{F}\xi _{N}/(\xi _{F}d_{N})$, $\gamma _{BM}=\gamma _{BN}d_{N}/\xi _{N}$.
The solutions of the above equations are%
\begin{equation}
\begin{array}{c}
\theta _{+}=4\arctan \left[ \tan (\frac{\theta _{0}}{4})\exp (-\frac{x}{\mu }%
)\right] , \\
\theta _{-}=\theta _{-\infty }+4\arctan \left[ \tan (\frac{\theta
_{0}-\theta _{-\infty }}{4})\exp (\frac{x}{\eta })\right] , \\
\theta _{-\infty }=\arctan \frac{\sin \theta _{S}}{\gamma _{BM}(\gamma k%
\widetilde{\Omega }+\Omega )+\cos \theta _{S}}, \\
\theta _{0}=2\arctan \frac{\sin \frac{\theta _{-\infty }}{2}}{\cos \frac{%
\theta _{-\infty }}{2}+\eta /\mu },%
\end{array}%
\end{equation}%
and normalized DOS at energy $\varepsilon $ is given by:
\begin{equation}
\nu = Re\left[ \cos \theta (-i\varepsilon + \delta )\right],   \label{eq:DoS}
\end{equation}
where $\delta=10^{-3}$ was used in calculations.

Fig.\ref{th} shows the results of calculation of total DOS (summed over both spin subbands)
from Eq.\ref{eq:DoS}. It is seen that the  peaks in DOS which for $x<0$
occur at energies $\varepsilon=\Delta,\,\varepsilon_{+},\,%
\varepsilon_{-}$ transform to dips for $x>0$.
The structure at energies $\varepsilon_{\pm}$ corresponds to minigap splitting due to effective exchange field $h_{ef} = E_{ex} \nu_F d_F/(\nu_F
d_F+\nu_N d_N)$, where $\nu_{N,F}, d_{N,F}$ are the normal-state densities
of states and thicknesses of N and F layers.
Interestingly, the double-peak structure at $\varepsilon_{\pm}$
at $x<0$ transforms to the double-dip structure in the bridge region ($x>0$)
at distances of the order of $\xi_N$.
The energy separation ($\varepsilon_{+}-\varepsilon_{-}$) between the peaks/dips
can be estimated as
$\varepsilon_{+}-\varepsilon_{-} \simeq \gamma_{BM} h_{ef}.$
For $E_{ex}=0,$ $h_{ef}$ is also zero and these features merge into a single peak (dip).

In the simplest approach, the resistance $(dU/dI)$ of SNF-NF-NFS structure
is determined by the renormalized diffusion coefficient in the NF bridge area
and the dips in DOS at $x>0$ should lead to the double-peak structure in $(dU/dI)$ vs $U$
similar to the observed experimentally (see Fig.\ref{result}(a)). Quantitative model
is beyond the frame of our model due to complex device geometry and a number of unknown parameters.

To conclude, we have observed experimentally a manifestation of the
superconducting minigap splitting in the N-layer contacted both with
superconductor and ferromagnet in complex planar S-(N/F)-S system formed by
Al-(Cu/Fe)-Al submicron-size bridge. Such a splitting has to exist in SF
bilayers also, but it is difficult to observe it there because of the large
values of the exchange field for conventional ferromagnets. It has been
demonstrated that the splitting occurs only for contacts to ferromagnetic
layers with uniform magnetization and disappears when the applied magnetic
field close to the coercive field. DOS calculations for SNF systems have
shown that the minigap splitting is really possible for parameters close to
experimental ones.

We acknowledge  A.V. Ustinov  for support and usefull discussions. The work was encouragement by grants of Russian Academy of Sciencies and Russian Foundation for Basic Research.


\begin{thebibliography}{99}
\bibitem{Gol_Kupr_1988} A.~A. Golubov, M.~Yu. Kupriyanov, Journal of Low
Temp. Phys. \textbf{70}, 83 (1988)

\bibitem{Gueron} S. Gueron, H. Pothier, N.~O. Birge, D. Esteve, and M.~H.
Devoret, Phys. Rev. Lett. \textbf{77}, 3025 (1996)

\bibitem{Courtois} A.~K. Gupta, L. Cretinon, N. Moussy, B. Pannetier, and H.
Courtois, Phys. Rev. B \textbf{69}, 104514 (2004)

\bibitem{Esteve} H.~le Sueur, P. Joyez, H. Pothier, C. Urbina, and D. Esteve,
Phys. Rev. Lett. \textbf{100}, 197002 (2008)

\bibitem{Giazotto} M. Meschke, J.~T. Peltonen, J.~P. Pekola, and F.
Giazotto, Cond-mat. arXiv:1105.387v1 (2011)

\bibitem{Cuevas} J.~C. Cuevas, J. Hammer, J. Kopu, J.~K. Viljas, and M.
Eschrig, Phys. Rev. B \textbf{73}, 184505 (2006)

\bibitem{SFIFS} A.~A. Golubov, M.~Yu. Kupriyanov, Ya.~V. Fominov, Pis'ma v ZhETF \textbf{75}, 223 (2002) [JETP
Lett., \textbf{75}, 190 (2002)]

\bibitem{Oboznov} V.~A. Oboznov, V.~V. Bol'ginov, A.~K. Feofanov, V.~V.
Ryazanov and A.~I. Buzdin, Phys. Rev. Lett. \textbf{96}, 197003 (2006)

\bibitem{Kontos} T. Kontos, M. Aprili, J. Lesueur, and X. Grison, Phys. Rev.
Lett. \textbf{86}, 304 (2001)

\bibitem{Yip} S.~-K. Yip, Phys. Rev. B \textbf{62}, R6127 (2000)

\bibitem{Karm1} T.~Yu. Karminskaya, M.~Yu. Kupriyanov, Pis'ma v ZhETF \textbf{85}, 343 (2007) [JETP
Lett. \textbf{85}, 286 (2007)]

\bibitem{Hanson} M. Hanson, O. Kazakova, P. Blomqvist, R. Wa\"ppling, and B. Nilsson, Phys. Rev. B \textbf{66}, 144419 (2002)

\bibitem{magnSim} W. Rave and A. Hubert, IEEE Trans. Magn. \textbf{36}, 3886 (2000)

\bibitem{BeckmannCAR}
D.~Beckmann, H.~B. Weber and H.v.~L\"oneysen, Phys. Rev. Lett. {\bfseries{93}}, 197003 (2004)

\bibitem{AMRexp} P.~P. Freitas and T.~S. Plaskett, J. Appl. Phys. \textbf{67}, 4901 (1990)

\bibitem{AMRt} R. McGuire, R.~J. Potter, IEEE Trans. Magn. \textbf{11}, 1018 (1975)

\bibitem{SNSTG}
T.~E. Golikova et al to be published

\bibitem{wsxm}
I. Horcas, R. Fernandez, J.~M. Gomez-Rodriguez, J. Colchero, J. Gomez-Herrero and A.~M. Baro, Rev. Sci. Instrum. \textbf{78} , 013705 (2007)


\bibitem{Kimura}
T. Kimura, T. Sato, and Y. Otani, Phys. Rev. Lett. \textbf{100}, 066602 (2008)

\bibitem{KL} M.~Yu. Kupriyanov and V.~F. Lukichev, Zh. Eksp. Teor. Fiz.
\textbf{94}, 139 (1988) {[}Sov. Phys. JETP \textbf{67}, 1163 (1988){]}.

\end{thebibliography}
\end{document}